\input harvmac
\input epsf
\def\lf{16\pi^2}
\def\llf{(16\pi^2)^2}
\def\lllf{(16\pi^2)^3}

\def\frak#1#2{{\textstyle{{#1}\over{#2}}}}
\def\frakk#1#2{{{#1}\over{#2}}}

\def \in{\leftskip = 40 pt\rightskip = 40pt}
\def \inn{\leftskip = 70 pt\rightskip = 70pt}
\def \out{\leftskip = 0 pt\rightskip = 0pt}

\def\yhat{\hat{y}}
\def\Bhat{\hat{B}}
\def\Khat{\hat{K}}

\def\sy{supersymmetry}
\def\sic{supersymmetric}

\def\cmp{Comm.\ Math.\ Phys.\ }

\def\ijmpa{{Int.\ J.\ Mod.\ Phys.\ }{\bf A}}

\def\npb{{Nucl.\ Phys.\ }{\bf B}}
\def\physrep{Phys.\ Reports\ }
\def\plb{{Phys.\ Lett.\ }{\bf B}}

\def\prd{{Phys.\ Rev.\ }{\bf D}}
\def\prl{Phys.\ Rev.\ Lett.\ }

\def\tmp{Theor.\ Math.\ Phys.\ }

{\nopagenumbers
\line{\hfil LTH 392}
\line{\hfil hep-ph/9705328}
\vskip .5in
\centerline{\titlefont $\beta$-functions in 
large-$N_f$ supersymmetric gauge theories}
\vskip 1in
\centerline{\bf P.M.~Ferreira, I.~Jack, D.R.T.~Jones and C.G.~North}
\bigskip
\centerline{\it Dept of Mathematical Sciences, 
University of Liverpool, Liverpool L69 3BX, U.K.}
\vskip .3in

We present calculations of the leading and $O(1/N_f)$ terms in a large-$N_f$
expansion of the $\beta$-functions and anomalous dimensions for various 
supersymmetric gauge theories, including supersymmetric QCD. In the case 
of supersymmetric QCD, we show that our $O(1/N_f)$ approximation displays
an infra-red fixed point in the conformal window 
$\frak{3}{2}N_c < N_f < 3N_c$. 

\Date{May 1997}

\newsec{Introduction}

The large-$N$ expansion is an alternative to conventional perturbation
theory. In both QCD and \sic\ QCD (SQCD), the large $N_c$ expansion is
of particular interest \ref\witt{ E. Witten,  Ann. Phys. 128 (1980)
363}; more tractable, however is the large $N_f$ expansion, and it is
this that we study here, for SQCD. In any theory, calculation correct to
some non-trivial order in $1/N$ requires the summation of one or more
infinite sets of Feynman diagrams, and   hence possible insight into the
large-order behaviour of perturbation theory.  There is a considerable
literature devoted to the study of large order perturbation
theory~\ref\zinn{J.~Zinn--Justin, \physrep 70 (1981) 110},
but remarkably little of it concerned with \sic\ theories. 
[For a discussion
of the large order behaviour of the \sic\ anharmonic oscillator, see
Ref.~\ref\verb{J.J.M. Verbaarschot and P. West, \prd 42 (1990) 1276;
{\it ibid\/}  {\bf D}43 (1991) 2718}; and for investigation of the role
of renormalons in various \sic\ theories see Ref.~\ref\verba{J.J.M.
Verbaarschot and P. West, \ijmpa 6 (1991) 2361}. 
Supersymmetric $\sigma$-models have been studied at large $N$ 
using critical methods
\ref\honk{A.N~Vasil'ev, Yu.M. Pis'mak and J.R. Honkonen,  \tmp 46 
(1981) 157; {\it ibid\/} 47 (1981) 291}  in a series of papers 
by Gracey: see for example Ref.~\ref\graceyx{J.A.~Gracey, \npb 352 (1991) 
183}.]
In this paper we calculate the leading and $O(1/N_f)$
terms in the gauge beta--function, $\beta_g$, 
for SQCD\foot{For QCD at large $N_f$
see \ref\graceya{J.A.~Gracey, \ijmpa 8 (1993) 2465;  
\plb 373 (1996) 173; \npb 414 (1994) 614}}. The type of bubble-sums we
confront are in fact similar to those in Ref.~\verba; the difference
being that they were concerned with renormalon singularities in
amplitudes, whereas, our interest being in beta--functions,    we
require the ultra-violet divergent terms from such sums. We find that
all our results can be expressed in terms of a simple function of the
coupling constant, and that although infinite classes of diagrams have
been summed, the resulting coefficient of $1/N_f$ has a finite radius of
convergence in $g$. Of course we cannot expect the expansion in
powers of $1/N_f$ to be convergent; it might well, however, be
Borel-summable, as has been 
conjectured\ref\bag{C.~Bagnuls and C.~Bervillier, hep-th/9702149}\ for the
expansion in $g$.

Some of our results have already appeared in Ref.~\ref\largen{   
P.M.~Ferreira, I.~Jack, and D.R.T.~Jones, hep-ph/9702304};
here we give more calculational detail and also extract the 
SQCD case. We use the superfield formalism, allied with supersymmetric 
dimensional regularisation and minimal subtraction 
(DRED)\ref\dred\ldf{W.~Siegel, \plb84 (1979) 193\semi
D.M.~Capper, D.R.T.~Jones and P.~van Nieuwenhuizen,
\npb167 (1980) 479}. 
As always with superfield perturbation theory, the calculation
of a typical contribution  consists of first reducing a superfield
Feynman diagram to a normal one by performing $D$-algebra, and
then performing the resulting Feynman integral. The first part is
straightforward, because our diagrams consist of bubble chains
grafted on to otherwise (at most) three-loop graphs. The second part
appears formidable; but fortunately the key to its performance
has been provided  by Palanques-Mestre  and
Pascual\ref\pmpp{A. Palanques-Mestre  and P.~Pascual, \cmp 95 (1984) 277},
who carried out similar non-\sic\ calculations in the abelian case.
The crucial realisation is that the  automatic 
cancellation of non-local counter-terms leads to apparently 
miraculous (but easily verified)
identities which simplify the summation over subtractions.

In the next section we carry out our program for an Abelian theory, 
with a superpotential selected so that ${\cal N} =2$ \sy\ is
included as a special case. Then in section~3 we show how with
very little extra work we can extract the corresponding result
for the non-Abelian case, by exploiting the fact that ${\cal N} =2$
theories are finite beyond one loop. In section~4 we show how the SQCD
result can be also deduced, and in section~5 we compare our result with 
perturbation theory in $g$ for the SQCD case. 

\newsec{General Abelian Theory}

Here we present results for a general theory produced by a $U_1$ gauging of 
the Wess-Zumino model with a superpotential given by
\eqn\none{
W = \frakk{\lambda}{\sqrt{N_f}}\sum_{i=1}^{N_f}\phi\xi_i\chi_i.}
Suppose we require a  
gauging such that the charges $q_{\chi_i}\equiv q_{\chi}$ and 
$q_{\xi_i}\equiv q_{\xi}$ 
are independent of $i$. Charge conservation and anomaly cancellation then
require 
\eqn\gatab{
q_{\phi}+q_{\chi}+q_{\xi}=
q^3_{\phi}+N_f(q^3_{\chi}+q^3_{\xi})=0,}
yielding
\eqn\gatac{
(N_f-1)(q^3_{\chi}+q^3_{\xi})-3(q_{\chi}^2q_{\xi}+q_{\chi}q_{\xi}^2)=0.}
An obvious solution is $q_{\xi}=-q_{\chi}$, $q_{\phi}=0$. In fact if this does
not hold, then dividing Eq.~\gatac\ by $q_{\chi}+q_{\xi}$ we obtain
\eqn\gatad{
(N_f-1)(q_{\chi}^2+q_{\xi}^2)-(N_f-2)q_{\chi}q_{\xi}=0,}
which is easily seen to have no solutions (except in the case $N_f=1$, with 
either $q_{\xi}=0$ or $q_{\chi}=0$, which are clearly not interesting from
our present large-$N_f$ perspective). So we now set 
$q_{\phi} = 0$ and $q_{\xi} = - q_{\chi} = q$; we will take 
$q=g/\sqrt{N_f}$. 
For the 
special case $\lambda = \sqrt{2}g$ we have 
${\cal N} =2$ \sy. In Fig.~1 we show the Feynman diagrams we require for 
$\beta_g$. 
\bigskip
\epsfysize= 2.0in
\centerline{\epsfbox{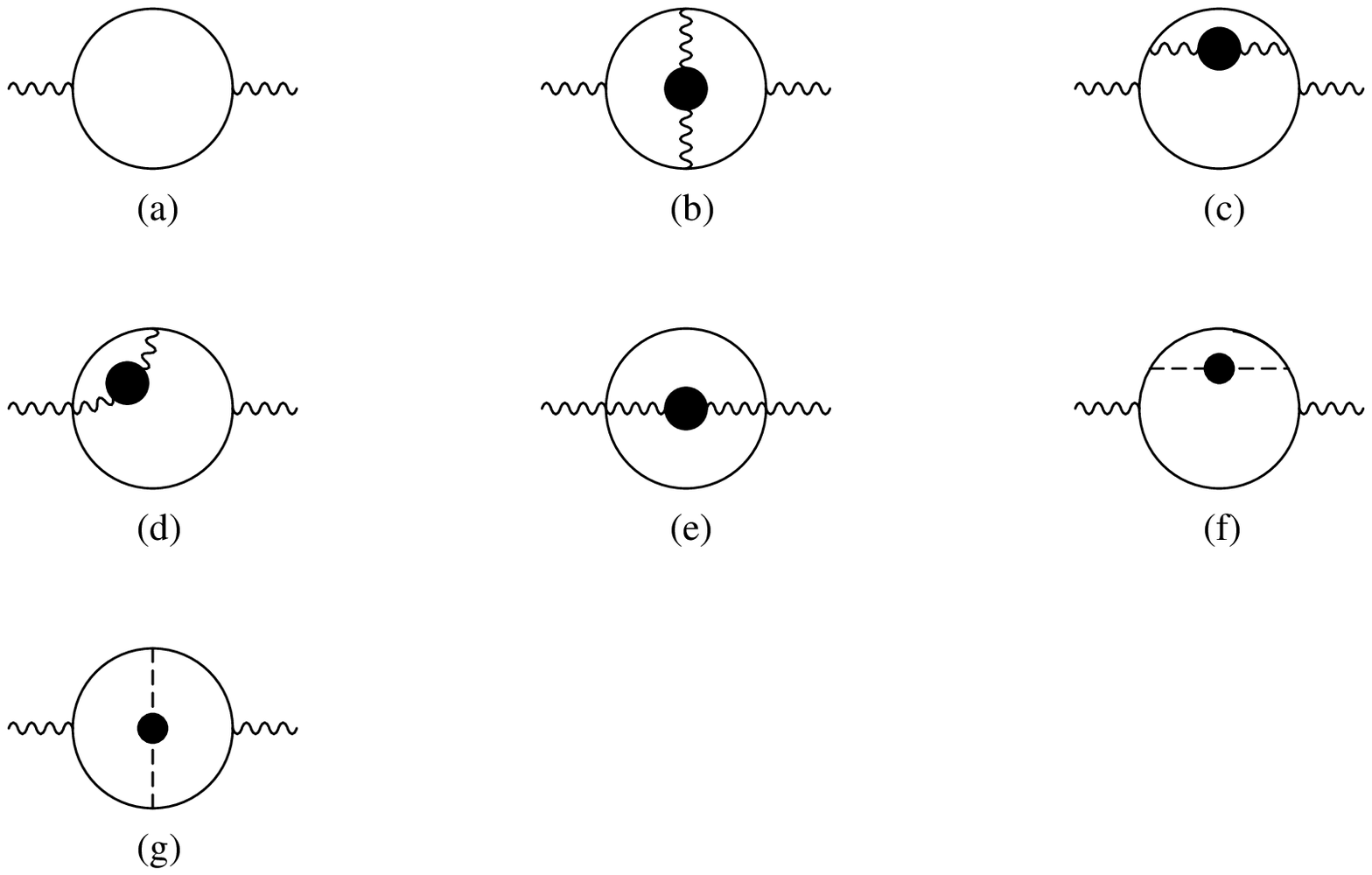}}
\in
{\it \noindent Fig.~1:
Feynman diagrams for section~2. Wavy lines are vector propagators,
solid lines are $\xi$ or $\chi$ propagators, and dashed lines are 
$\phi$ propagators. Blobs denote sums of chains
of $\xi, \chi$ bubbles.}
\bigskip
\out
A blob on a diagram represents the sum of chains of $\xi, \chi$ bubbles of 
arbitrary length, as shown in Fig.~2(a) (for a $\phi$-propagator) and Fig.~2(b)
(for a vector propagator). We
choose to work in the Landau gauge; this means that the diagram with no
bubble on a vector propagator can be subsumed with the diagrams with chains
of bubbles, since a chain of one or more bubbles produces a tranverse projection
operator. Fig.~1(a) is $O(1)$, while Figs.~1(b)-(g) are all $O(1/N_f)$.   
\bigskip
\epsfysize= 1in
\centerline{\epsfbox{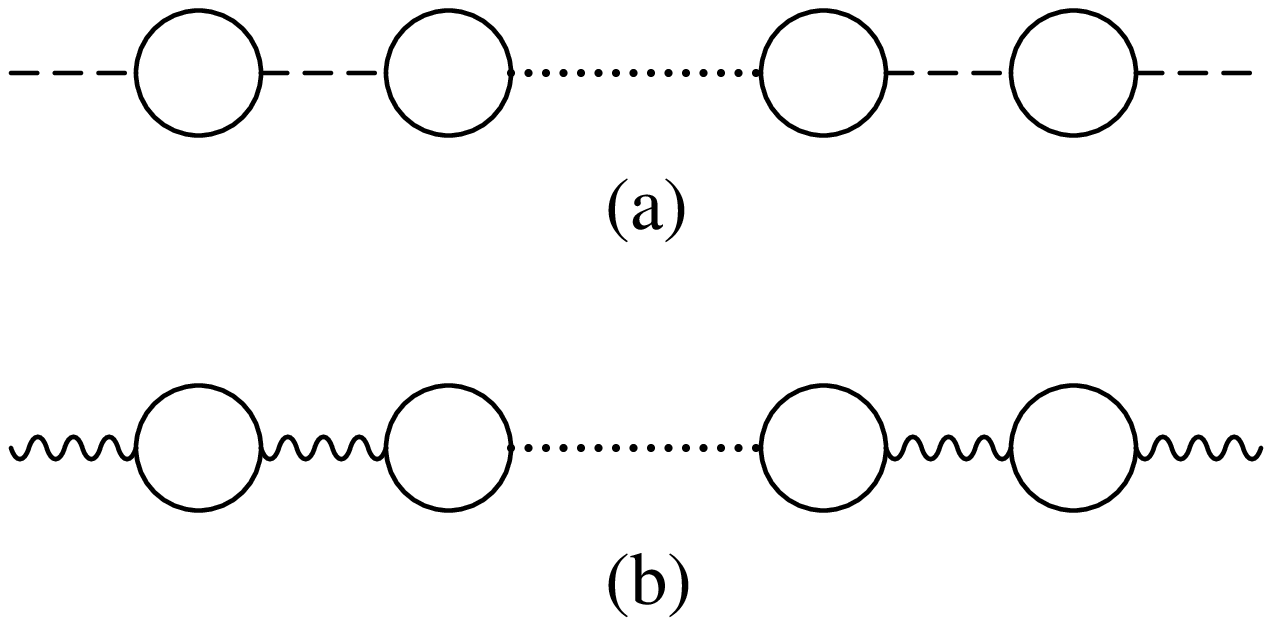}}
\inn
{\it \noindent Fig.~2:
Bubble chains on chiral and vector propagators.}
\bigskip
\out
For details of our technique for dealing with the $D$-algebra part
of the calculation we refer the reader to 
Ref.~\ref\cjjb{I.~Jack, D.R.T.~Jones and C.G.~North, \npb 486 (1997) 479}, 
where we found $\beta_g$ for
an abelian theory to four loops by calculating the
vector superfield self-energy. The upshot is that
each diagram with a blob 
reduces to one of a basic set of Feynman integrals, depicted in
Fig.~3. Again in Fig.~3,
a chain of bubbles is understood to represent a summation over
arbitrary numbers of bubbles.
\bigskip
\epsfysize= 1.0in
\centerline{\epsfbox{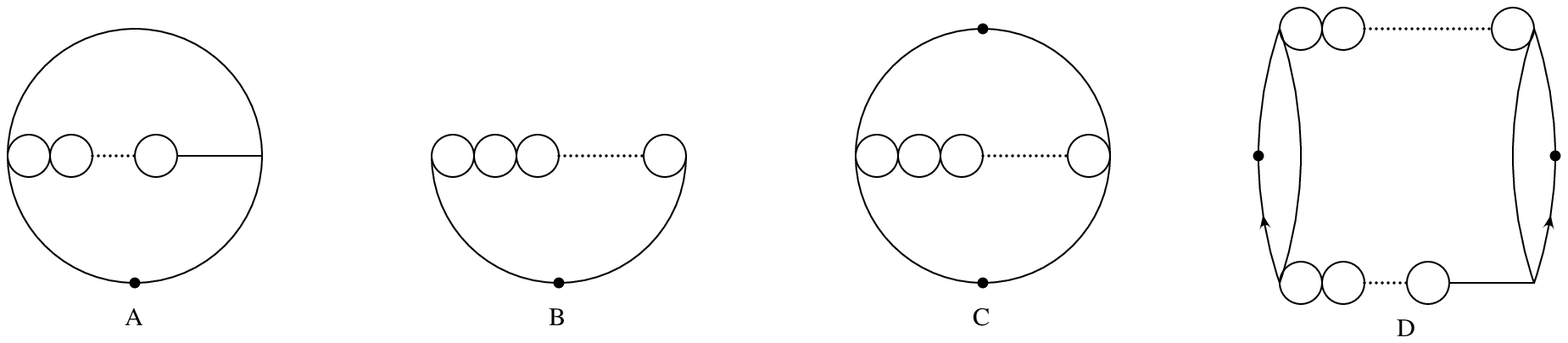}}
\in
{\it \noindent Fig.~3:
Feynman diagrams representing the bubble sums $A, B$ $C$ and $D$. The black
dots denote squared propagators and pairs of arrows denote contracted momenta.}
\bigskip
\out
The Feynman integrals (A)-(D) are evaluated in Appendix~A, with results
given in Eqs.~(A.17), (A.8), (A.18) and (A.25) respectively. 
We note all the bubble sums relevant to our calculations depend on the function
$G(x)$, which has a zero at $x=1$ and a simple pole at $x=\frak{3}{2}$.
We may therefore anticipate that our results in this and subsequent sections
will have a finite radius of convergence in the appropriate coupling
constant, because of this pole.
The contributions 
from the diagrams in Fig.~1 are given in terms of the 
diagrams of Fig.~3 in Table~1. In this Table and the following ones, the second
column represents the result of performing the $D$-algebra, and should be 
multiplied by the corresponding factor in the third column which represents
the symmetry factor and also (in the non-abelian case) products of group
matrices. An additional factor of $1/N_f$ is also 
understood in each case, and we have introduced $K=g^2/(8\pi^2)$ and 
$y=\lambda^2 / (16\pi^2)$. For ${\cal N} = 2$ \sy\ we have $y=K$. 
\medskip 
$$\vbox{\offinterlineskip
\def\vr{\vrule height 11pt depth 5pt}
\def\vrq{\vr\quad}
\settabs
\+
\vrq Diagram  \quad & \vrq \qquad Bubble Sum
 \quad\qquad\
& \vrq \quad$16X_3+32g^6\tr[C(R)^3]$\quad&
\vr\cr\hrule
\+
\vrq Diagram \quad & \vrq \quad
 Bubble Sum\quad\quad
& \vrq Factor&
\vr\cr\hrule
\+
\vrq 1(b)\quad & \vrq  \quad ${1\over2}C(K)-2A(K)$
& \vrq \quad$2gK^2$&
\vr\cr\hrule 
\+
\vrq 1(c)\quad & \vrq  \quad ${1\over2}[{1\over K}B(K)-A(K)]$              
& \vrq \quad$4gK^2$&          
\vr\cr\hrule
\+
\vrq 1(d)\quad & \vrq  \quad ${1\over2}[{1\over K}B(K)-A(K)]$              
& \vrq \quad$-8gK^2$&          
\vr\cr\hrule
\+
\vrq 1(e)\quad & \vrq  \quad ${1\over K}B(K)$              
& \vrq \quad$2gK^2$&          
\vr\cr\hrule
\+
\vrq 1(f)\quad & \vrq  \quad $A(y)$              
& \vrq \quad$2gKy$&          
\vr\cr\hrule
\+
\vrq 1(g)\quad & \vrq  \quad $C(y)$              
& \vrq \quad$-gKy$&          
\vr\cr\hrule
}$$
\in
{\it \noindent Table~1: $O(1/N_f)$ contributions to $\beta_g$ for the abelian 
theory.}
\bigskip
\out

Figs.~1(b)-1(g) evidently have precise counterparts in a three-loop 
calculation with the blob representing just one bubble; and in fact  
the bubble-summed diagrams (A)-(D) of Fig.~3 are generalised forms of 
those in Fig~2(A)-(D) of Ref.~\cjjb. The results in Table~1 for 
Figs.~1(b)-1(g) can hence be 
read off from the corresponding results of Table~1 in Ref.~\cjjb. 
Using our results from Appendix A for Fig.~3(A)-(D) and 
including the $O(1)$ contribution 
of Fig.~1(a), we obtain 
\eqn\gata{
\beta_g = gK\left[1 + \frakk{2}{N_f}
\int^{K}_{y} (1-2x)G(x)\,dx\right].}
The anomalous dimensions through $O(1/N_f)$ are given by Fig.~4.
\bigskip
\epsfysize= 2.0in
\centerline{\epsfbox{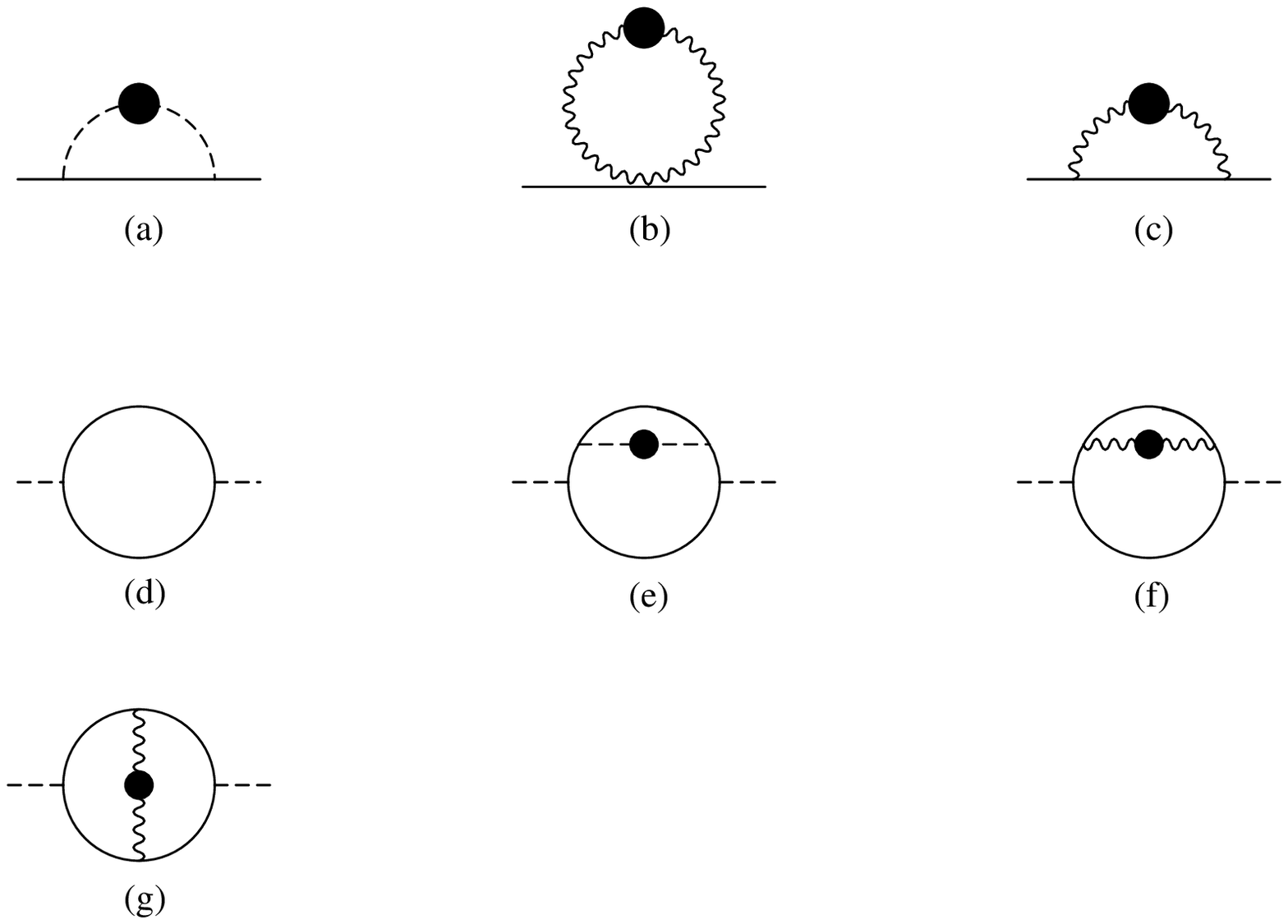}}
\in
{\it \noindent Fig.~4:
Feynman diagrams for the anomalous dimensions }
\medskip
\out
 The contributions from these diagrams are given in Table~2,
in terms of the basic diagrams of Fig.~3.
$$\vbox{\offinterlineskip
\def\vr{\vrule height 11pt depth 5pt}
\def\vrq{\vr\quad}
\settabs
\+
\vrq Diagram  \quad & \vrq \qquad Bubble Sum
 \quad\qquad &\vrq \qquad Factor \qquad&
\vr\cr\hrule
\+
\vrq Diagram \quad & \vrq \quad
 Bubble Sum\quad\quad&\vrq\qquad Factor \qquad
\vr\cr\hrule
\+
\vrq 4(a)\quad & \vrq  \quad $B(y)$ &\vrq$y$ &
\vr\cr\hrule 
\+
\vrq 4(b)\quad & \vrq  \quad $B(K)$ &\vrq$-K$ &                   
\vr\cr\hrule                 
\+ 
\vrq 4(c)\quad & \vrq  \quad $0$ &\vrq &
\vr\cr\hrule
\+
\vrq 4(e)\quad & \vrq  \quad $A(y)$ &\vrq$2y^2$ &                  
\vr\cr\hrule                 
\+
\vrq 4(f)\quad & \vrq  \quad ${1\over K}[B(K)-1]-A(K)$ &\vrq $2yK$ & 
\vr\cr\hrule                 
\+
\vrq 4(g)\quad & \vrq  \quad ${1\over K}[B(K)-1]$ & \vrq$-2yK$ & 
\vr\cr\hrule                 
}$$
\in
{\it \noindent Table~2: $O(1/N_f)$ contributions to anomalous dimensions for
the abelian theory}
\medskip
\out
The contributions from graphs with one or more bubbles in Fig.~4(c) are
guaranteed to be zero since the bubbles lead to a transverse projection 
operator on the vector propagator, giving zero when attached to an external 
line. Since we have chosen the Landau gauge, the zero-bubble contribution to 
Fig.~4(c) will also give zero by the same token. The same reasoning will be 
used later to deduce 
that Figs.~6(b),(d) give zero.  
Adding the contributions from Figs.~4(a)-(c), and those from Figs.~4(d)-(g),
we obtain
\eqn\gatd{\eqalign{
\gamma_{\xi}=\gamma_{\chi}&=\frakk{1}{N_f}\left[yG(y)-KG(K)\right],\cr
\gamma_{\phi}&= y+\frakk{2y}{N_f}\left[G(K)-G(y)+2\int_y^{K}G(x)\, dx
\right].\cr}}

It is easy to verify that our result for $\beta_g$ in Eq.~\gata\ reproduces the 
relevant terms in  
the three and four 
loop calculations presented in Ref.~\cjjb. The results for $\gamma_{\xi,\chi}$
and $\gamma_{\phi}$ in Eq.~\gatd\ agree with the three-loop results of 
Ref.~\ref\cjjc{I.~Jack, D.R.T~Jones and C.G.~North, 
\npb 473 (1996) 308}, and in the ungauged case agree with the four-loop results
of Ref.~\ref\fjj{P.M.~Ferreira, I.~Jack, and D.R.T.~Jones, \plb 392 (1997) 
376}\ for a
generalised Wess-Zumino model.
Moreover, for ${\cal N}=2$ (which 
corresponds to $y = K$) we have $\beta_g = \gamma_{\phi}= 0$ beyond 
one loop, 
and $\gamma_{\xi} = \gamma_{\chi} = 0$ to all orders, in accordance 
with 
Ref.~\ref\hsw{
P.S.~Howe, K.S.~Stelle and P.~West, \plb 124 (1983) 55\semi
P.S.~Howe, K.S.~Stelle and P.K.~Townsend, \npb  236 (1984) 125}.
The results of Eqs.~\gata, \gatd\ may readily be specialised to the case of 
\sic\ QED simply by setting $y=0$. 
  
\newsec{General Non-Abelian Theory}

We now consider a non-abelian theory with gauge group ${\cal G}$ and 
superpotential \eqn\nata{W=\frakk{\lambda}{\sqrt
N_f}\phi^a\sum_{i=1}^{N_f}\xi_i^TS_a\chi_i,} where $\xi_i, \chi_i, \phi$
are multiplets transforming under the  $S^*$, $S$ and adjoint
representations of ${\cal G}$ respectively. For  notational simplicity
we  take the representation $S$ to be irreducible for the time being; 
we shall present the results for a reducible representation in due
course.    In addition to diagrams similar in form to those computed
earlier in the abelian case and  shown in Fig.~1, the two-point function
for the vector superfield  includes the additional diagrams depicted in
Fig.~5, because the  $\phi$ field now has gauge interactions. There are
also further  diagrams involving the gauge coupling $g$ only, which we
shall be able to avoid computing.  

The diagrams of Figs.~1(f),(g) and Fig.~5 contain no internal vectors, 
and produce contributions to $\beta_g$ which contain the
Yukawa coupling $y$ (apart from the zero-bubble contribution to
Fig.~5(a)).  They are in fact the only graphs which contribute $y$
dependent terms up to $O(1/N_f)$. Graphs with no vector propagators will
be the same in the background field gauge as in an  ordinary gauge. Now
in the background field gauge, $\beta_g$  is given by  the
vector-field two-point function even in the non-abelian case. We deduce 
that the contribution to $\beta_g$ at $O(1/N_f)$ which
contains $y$  is correctly given by Figs.~1(f),(g) and Fig.~5.    Upon
performing the $D$-algebra, these diagrams all  give rise to Feynman
integrals with  bubble sums as depicted in Fig.~3.  
The contributions from the diagrams of the form Figs.~1(f),(g) and Fig.~5 are 
listed in Table~3. 
\bigskip
\epsfysize= 1.0in
\centerline{\epsfbox{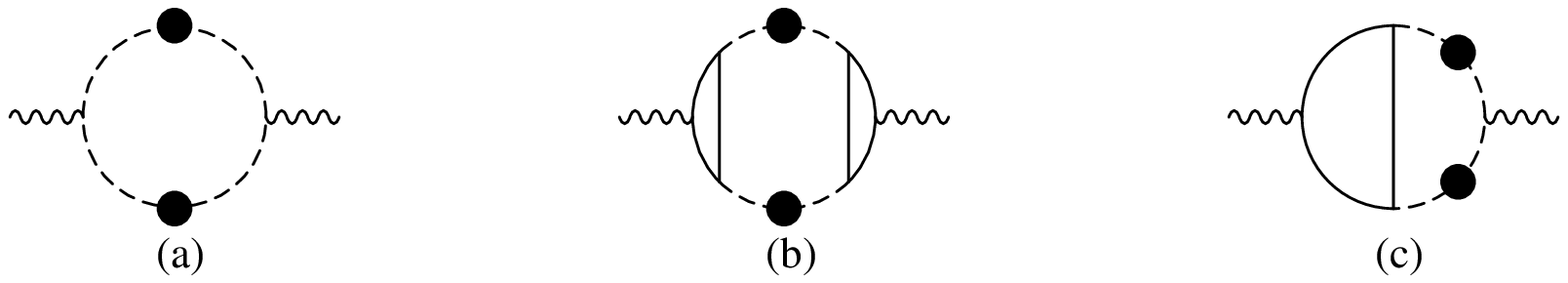}}
\in
{\it \noindent Fig.~5:
Additional vector two-point diagrams for the non-abelian theory.}

\medskip
\out
$$\vbox{\offinterlineskip
\def\vr{\vrule height 11pt depth 5pt}
\def\vrq{\vr\quad}
\settabs
\+
\vrq Diagram  \quad & \vrq \qquad Bubble Sum
 \quad\qquad\
& \vrq \quad$16X_3+32g^6\tr[C(R)^3]$\quad&
\vr\cr\hrule
\+    
\vrq Diagram \quad & \vrq \quad 
 Bubble Sum\quad\quad
& \vrq Group factor&
\vr\cr\hrule
\+
\vrq 1(f)\quad & \vrq  \quad $A(\yhat)$
& \vrq \quad$2gK\yhat C_S$&
\vr\cr\hrule
\+
\vrq 1(g)\quad & \vrq  \quad $C(\yhat)$
& \vrq \quad$-gK\yhat[C_S-{1\over2}C(G)]$&
\vr\cr\hrule
\+
\vrq 5(a)\quad & \vrq  \quad $\Bhat (\yhat)$
& \vrq \quad${1\over2}gKC(G)$&
\vr\cr\hrule
\+
\vrq 5(b)\quad & \vrq  \quad $-{\tilde B}(\yhat)-2D(\yhat)$
& \vrq \quad${1\over2}gK\yhat^2C(G)$&
\vr\cr\hrule
\+
\vrq 5(c)\quad & \vrq  \quad $0$
& \vrq \quad&
\vr\cr\hrule
}$$
\in
{\it \noindent
Table~3: $O(1/N_f)$ contributions to $\beta_g$ in the non-abelian theory. }
\bigskip
\out
The second column of Table~3 requires some explanation. Firstly, we have 
defined $\yhat=yT(S)$, where $T(S)\delta_{ab}=\tr(S_aS_b)$. Secondly,
care needs to be taken with the graphs of Fig.~5, which contain two 
bubble-chains. This leads to extra combinatorial factors. In fact,
upon performing the $D$-algebra, 
Fig. 5(a) yields a Feynman integral with the topology of Fig.~3(B), but 
the calculation differs from that in Appendix A, in that Eq.~(A.8) 
is replaced 
by 
\eqn\nataa{
B(\kappa)\to \Bhat (\kappa)=\sum_{n=0}^{\infty}(n+1)G_n\kappa^n
={d\over{d\kappa}}[\kappa G(\kappa)].}
Similarly, Fig.~5(b) yields two Feynman integrals. 
One again has the topology 
of Fig.~3(B), but now with 
\eqn\natab{
B(\kappa)\to {\tilde B}(\kappa)=\sum_{n=0}^{\infty}(n+1)G_{n+2}\kappa^n
={d\over{d\kappa}}\left\{{1\over{\kappa}}
[G(\kappa)-1]\right\}.}
The other is depicted in Fig.~3(D), and yields a bubble-sum contribution given 
in Eq.~(A.25).
In the third column, we have defined $C(S)=C_S1$, where $C(S)=S_aS_a$, and used
${\tr[C(S)^2]\over {rT(S)}}=C_S$ for an irreducible representation (where $r$
is the number of group generators).  
Adding the contributions in Table~3, we find 
\eqn\nata{
-\frakk{2gKC_S}{N_f}
\int^{\hat y}_0(1-2x)G(x)\,dx
+\frakk{gK}{N_f}\left[{1\over2}-\int^{\hat y}_0G(x)\,dx \right]C(G).}
As argued earlier, this represents the full $y$-dependent contribution at
$O(1/N_f)$, together with a single $y$-independent term which represents the
(one-loop) contribution from the zero-bubble part of Fig.~5(a).
We can then infer the full non-abelian result by using the afore-mentioned 
fact that there are no divergences 
beyond one loop  for ${\cal N}=2$, i.e. when we set $y=K$, together with the 
fact that the part of $\beta_g$ we have not yet calculated is purely a 
function of $g$. We also need to include the rest of the 
one-loop $\beta_g$.
The result is

\eqn\natb{\eqalign{\beta_g=&gK\left[T(S)+\frakk{2C_S}{N_f}
\int_{\hat y}^{\hat K}
(1-2x)G(x)\,dx\right]
\cr
\qquad&+\frakk{gK}{N_f}\left[\int_{\hat y}^{\Khat}G(x)\,dx -1
\right]C(G),\cr}}
where $\Khat=KT(S)$.

The chiral superfield anomalous dimensions are given by diagrams of the form
of Fig.~4 together with the additional diagrams of Fig.~6. 
\bigskip
\epsfysize= 2in
\centerline{\epsfbox{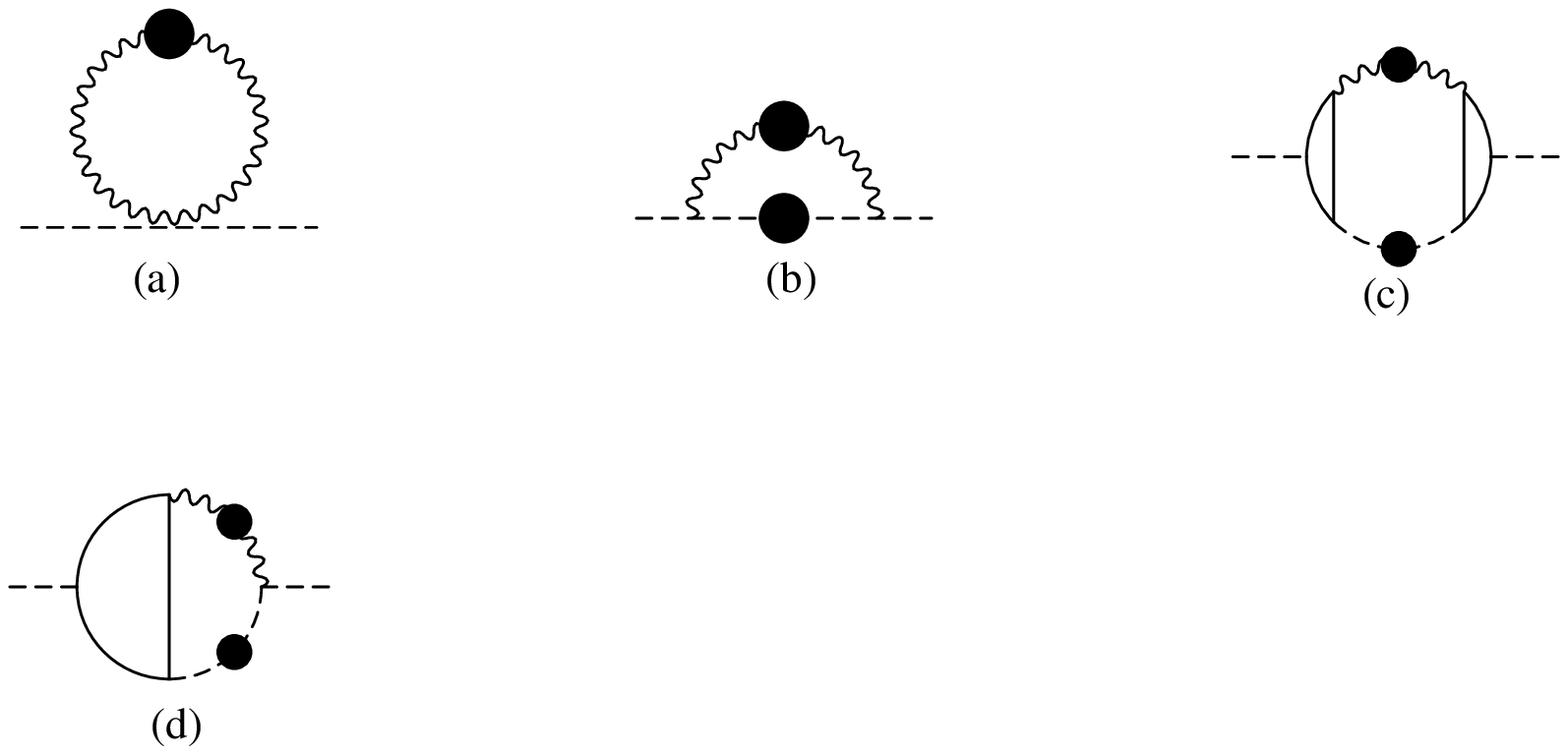}}
\in
{\it \noindent Fig.~6:
The additional Feynman diagrams for the anomalous dimension in the 
non-abelian theory. }
\medskip
\out
The individual contributions at $O(1/N_f)$ are given in Table~4.
$$\vbox{\offinterlineskip
\def\vr{\vrule height 11pt depth 5pt}
\def\vrq{\vr\quad}
\settabs
\+
\vrq Diagram  \quad & \vrq \qquad Bubble Sum
 \quad\qquad\
& \vrq \quad$16X_3+32g^6\tr[C(R)^3]$\quad&
\vr\cr\hrule
\+
\vrq Diagram \quad & \vrq \quad
 Bubble Sum\quad\quad
& \vrq Group factor&
\vr\cr\hrule
\+
\vrq 4(a)\quad & \vrq  \quad $B(\yhat)$
& \vrq \quad$yC(S)$&
\vr\cr\hrule
\+
\vrq 4(b)\quad & \vrq  \quad $B(\Khat)$
& \vrq \quad$-KC(S)$&
\vr\cr\hrule
\+ 
\vrq 4(c)\quad & \vrq  \quad $0$
& \vrq \quad&
\vr\cr\hrule
\+
\vrq 4(e)\quad & \vrq  \quad $A(\yhat)$
& \vrq \quad$2y\yhat C_S$&
\vr\cr\hrule 
\+
\vrq 4(f)\quad & \vrq  \quad ${1\over \Khat}[B(\Khat)-1]-A(\Khat)$             
& \vrq \quad$2y\Khat C_S$&        
\vr\cr\hrule
\+
\vrq 4(g)\quad & \vrq  \quad ${1\over \Khat}[B(\Khat)-1]$
& \vrq \quad$-2y\Khat[C_S-{1\over2}C(G)]$&
\vr\cr\hrule
\+
\vrq 6(a)\quad & \vrq  \quad $B(\Khat)$
& \vrq \quad$-KC(G)$&
\vr\cr\hrule   
\+
\vrq 6(b)\quad & \vrq  \quad $0$ 
& \vrq \quad&
\vr\cr\hrule   
\+
\vrq 6(c)\quad & \vrq  \quad $0$
& \vrq \quad&
\vr\cr\hrule
\+
\vrq 6(d)\quad & \vrq  \quad $0$   
& \vrq \quad&   
\vr\cr\hrule
}$$
\in
{\it \noindent
Table~4: $O(1/N_f)$ contributions to 
anomalous dimensions for the non-abelian 
theory.}
\bigskip
\out
The anomalous dimensions for $\xi$ or $\chi$ are given at $O(1/N_f)$ by 
adding the
contributions from Figs.~4(a)-(c), while the anomalous dimension for
$\phi$ is given up to $O(1/N_f)$ by adding the $O(1/N_f)$ contributions from 
Figs.~4(e)-(g), and  Figs.~6(a)-(c), together with the $O(1)$ 
contribution from Fig.~4(d). We obtain: 
\eqna\natc
$$\eqalignno{
\gamma_{\xi} &= \gamma_{\chi}
= \frakk{1}{N_f}\left[yG(\hat y)-KG(\hat K)\right]C(S),&\natc a\cr
\gamma_{\phi} &= \hat y+\frakk{2yC_S}{N_f}\left[G(\hat K)-G(\hat y)+
2\int_{\hat y}^{\hat K}G(x)\,dx\right]\cr
&+\frakk{1}{N_f}(y - K) G(\hat K) C(G)-\frakk{y}{N_f}C(G).&\natc b\cr}
$$
The above  results contain as special cases all those presented in 
previous sections. Once again one can check compatibility with the three 
and four-loop calculations from Ref.~\cjjb\ and 
Ref.~\cjjc. 

For completeness we present results for the case when $\chi, \xi$ 
transform according to a {\it reducible\/} representation. Suppose there are 
$n_{\alpha}$ pairs of multiplets $\xi^{\alpha}_{i_{\alpha}}$, 
$\chi^{\alpha}_{i_{\alpha}}$, each  
transforming according to a representation $S^{\alpha}$, and with
$\sum_{\alpha} n_{\alpha}=N_f$.   
The superpotential becomes:
\eqn\irra{W=\sum_{\alpha}\sum_{i_{\alpha}=1}^{n_{\alpha}}
\frakk{\lambda_{\alpha}}{\sqrt N_f}
\phi^a\xi^{\alpha}_{i_{\alpha}}{}^T S^{\alpha}_a\chi^{\alpha}_{i_{\alpha}},}
where $\alpha$ labels the irreducible representations. 
It is convenient to generalise the definitions of $\yhat$ and $\Khat$ 
as follows:
\eqn\nhats{\eqalign{
\yhat &= \frakk{1}{N_f}\sum_{\alpha}n_{\alpha}y_{\alpha}T(S_{\alpha})\cr
\Khat &= \frakk{g^2}{8\pi^2 N_f}\sum_{\alpha}n_{\alpha}T(S_{\alpha})\cr
}}
and introduce 
\eqn\deltas{\eqalign{
\Delta_K &= 
\frakk{K}{rN_f}\sum_{\alpha}n_{\alpha}\Tr\left[ C(S_{\alpha})^2\right]\cr
\Delta_y &= 
\frakk{1}{rN_f}\sum_{\alpha}n_{\alpha}y_{\alpha}
\Tr\left[ C(S_{\alpha})^2\right]\cr
\Delta_{y^2} &= 
\frakk{1}{rN_f}\sum_{\alpha}n_{\alpha}y_{\alpha}^2
\Tr\left[ C(S_{\alpha})^2\right].\cr}}

Our results are:

\eqn\deltasb{\eqalign{
{\beta_g\over g}&=\Khat+{2K\Delta_K\over {N_f\Khat}}\int_0^{\Khat}(1-2x)G(x)\,dx
-{2K\Delta_y\over{N_f\yhat}}\int_0^{\yhat}(1-2x)G(x)\,dx \cr
&\quad +{K\over N_f}C(G)\left[\int_{\yhat}^{\Khat}G(x)\,dx-1\right],\cr
\gamma_{\xi_{\alpha}} &= \gamma_{\chi_{\alpha}} 
= \frakk{1}{N_f}\left[y_{\alpha}G(\hat y)-KG(\hat K)\right]C(S_{\alpha}),\cr
\gamma_{\phi} &= \yhat 
+ \frakk{2\Delta_y K}{N_f\Khat}\left[G(\hat K)- 1 +
2\int_{0}^{\hat K}G(x)\,dx\right]\cr
&-\frakk{2\Delta_{y^2}}{N_f\yhat}\left[G(\yhat)-1 
+ 2\int_{0}^{\yhat}G(x)\,dx\right]\cr
&+\frakk{K\yhat}{N_f\Khat}C(G)\left[G(\Khat)-1\right]
-\frakk{K}{N_f}G(\Khat)C(G).\cr}}

\newsec{Supersymmetric QCD}
We can now deduce $\beta_g$ and the chiral superfield
anomalous dimension for large-$N_f$ supersymmetric QCD from the results of the 
previous section. We consider supersymmetric QCD with $N_f$ pairs of chiral
superfields $\xi_i$, $\chi_i$, transforming under the $S^*$, $S$ 
representations of ${\cal G}$ respectively. 
In principle, to obtain the results for supersymmetric
QCD from those for our general non-abelian theory, we need to remove 
the contributions of all diagrams involving the adjoint field $\phi$.
For the anomalous dimension of the chiral field, we simply need 
to remove the diagram Fig.~4(a). The result is then
\eqn\qcda{
\gamma_{\xi} = \gamma_{\chi}
= -\frakk{1}{N_f}KG(\hat K)C(S).}
We note that we may obtain this result simply by setting $y=0$ in 
Eq.~\natc{a}. 
In fact, the same principle may also be applied in the case of $\beta_g$. 
One can convince oneself that the
vector-field two-point diagrams with a $\phi$-propagator 
which contribute at $O(1/N_f)$ are precisely those of Fig.~1(f),(g) and Fig.~5.
These all contain $y$ except for the zero-bubble contribution to Fig.~5(a).
Hence, we can delete the contributions
of diagrams with adjoint fields simply by setting $y=0$ in Eq.~\natb,
provided we also subtract the contribution of the zero-bubble contribution to
Fig.~5(a) (i.e. the $y$-independent term in Eq.~\nata) by hand.
The result is
\eqn\qcdb{\eqalign{\beta_g=&gK\left[T(S)+\frakk{2C_S}{N_f}
\int_{0}^{\hat K} 
(1-2x)G(x)\,dx\right]
\cr
\qquad&+\frakk{gK}{N_f}\left[\int_{0}^{\hat K}G(x)\,dx -\frakk{3}{2}
\right]C(G).\cr}}
For the case of $SU(N_c)$ with $N_f$ flavours we find
\eqn\sqcda{\eqalign{\beta_g=
&\half gK -\frakk{3N_c}{2N_f}gK
+2gK\frakk{N_c}{N_f}\int_{0}^{\hat K} 
(1-x)G(x)\,dx
\cr
\qquad&-\frakk{gK}{N_c N_f}\int_{0}^{\hat K}(1-2x)G(x)\,dx .\cr}}
Note that in this case $\Khat = K/2$.

It is also of interest to consider the case of supersymmetric QCD coupled 
to a singlet $\phi$ by a superpotential
\eqn\sqcdb{
W=\frakk{\lambda}{\sqrt N_f}\phi\sum\xi_i^T\chi_i.}
The additional diagrams contributing to the vector two-point function are of the
form of Figs.~1(f),(g). Now these diagrams contain no internal vectors, and
hence would be the same in the background field gauge as in a conventional
gauge. But in the background gauge, $\beta_g$ 
is determined by the vector two-point diagrams. Hence
the additional contribution to $\beta_g$ can be computed 
from these diagrams. The contributions can be read off from the corresponding
entries in Table~1 by replacing $K$ by $\Khat$ in the last column. 
Hence $\beta_g$ is now given by
\eqn\sqcdc{\eqalign{\beta_g=&g\Khat\left[1-\frakk{2}{N_f}
\int_{0}^{y}(1-2x)G(x)\,dx\right]+\frakk{2gKC_S}{N_f}   
\int_{0}^{\Khat}(1-2x)G(x)\,dx\cr
&+\frakk{gK}{N_f}\left[\int_{0}^{\hat K}G(x)\,dx -\frakk{3}{2}
\right]C(G).\cr}}
The anomalous dimensions are now given by diagrams of the form Fig.~4.
The contributions from Figs.~4(a)-(c) can be read off from the corresponding
entries in Table~2 by replacing $K$ by $\Khat$ in the second column and by 
$KC(S)$ in the third column, giving
\eqn\sqcdd{
\gamma_{\xi}=\gamma_{\chi}=\frakk{1}{N_f}\left[yG(y)-KG(\Khat)C(S)\right]} 
The contributions from Figs.~4(e)-(g) can be read off from Table~2 by
replacing $K$ by $\Khat$ everywhere, giving
\eqn\sqcde{
\gamma_{\phi}= y+\frakk{2y}{N_f}\left[G(\Khat)-G(y)+2\int_y^{\Khat}G(x)\, dx
\right].}
For the case of $SU(N_c)$ with $N_f$ flavours we find
\eqn\sqcdf{\eqalign{\beta_g=  
&\half gK -\frakk{3N_c}{2N_f}gK
+2gK\frakk{N_c}{N_f}\int_{0}^{\hat K}
(1-x)G(x)\,dx
\cr
\qquad&-\frakk{gK}{N_c N_f}\int_{0}^{\hat K}(1-2x)G(x)\,dx 
-{gK\over{N_f}}\int_0^{y}(1-2x)G(x)\,dx.}}

\newsec{Infra-red fixed points}

In this section we compare our result for \sic\ QCD with 
perturbation theory in the gauge coupling. We begin by giving the 
known results at one through three loops:
\foot{Recall that our gauge coupling is $g/\sqrt{N_f}$}
\eqn\loopa{\eqalign{
\lf\beta_g^{(1)} &= \left(1 - 3{N_c\over{N_f}}\right)g^3,\cr
\llf\beta_g^{(2)} &= \left(\left[4N_c-{2\over {N_c}}\right]{1\over{N_f}}  
-6{N_c^2\over{N_f^2}}\right)g^5,\cr
\lllf\beta_g^{(3)} &= \left(\left[{3\over{N_c}}-4N_c\right]{1\over{N_f}}
+\left[21N_c^2-{2\over{N_c^2}}-9\right]{1\over{N_f^2}}
-21{N_c^3\over{N_f^3}} 
\right)g^7. }}
For $\beta_g^{(4)}$ we have only a partial result\cjjc: 
\eqn\loopb{\eqalign{
(16\pi^2)^4\beta_g^{(4)}=&\biggl(-\frakk{2}{3}\frakk{1}{N_cN_f}
+\Bigl[-\bigl(\frakk{62}{3}+2\kappa+8\alpha\bigr)N_c^2+\frakk{100}{3}
+4\alpha+\frakk{6\kappa-20}{3N_c^2}\Bigr]{1\over{N_f^2}}\cr
&+\Bigl[36(1+\alpha)N_c^3-(34+12\alpha)N_c-\frakk{8}{N_c}
-\frakk{4}{N_c^3}\Bigl]{1\over{N_f^3}}-(6+36\alpha){N_c^4\over{N_f^4}}
\biggr)g^9\cr}}
where $\alpha$ is an as yet undetermined parameter, and where 
$\kappa=6\zeta(3)$. 

It is easy to show\ref\timb{D.R.T~Jones, \npb 87 (1975) 127} 
that $\beta_g^{(2)} > 0$ when $\beta_g^{(1)}=0$. For 
$\beta_g^{(1)}$ less than but near zero, it follows that there exists 
an infra-red fixed point in the evolution of $g$ under 
renormalisation~\foot{This holds also for non-supersymmetric QCD
\ref\cas{A.~Belavin and A.~Migdal, JETP Lett. 19 (1974) 181\semi
W.~Caswell, \prl 33 (1974) 244\semi 
D.R.T.~Jones, Proceedings, Recent Progress In Lagrangian Field Theory and
Applications, Marseille 1974, 68}}. 
According to Seiberg\ref\seib{N.~Seiberg, \npb 435 (1995) 129}\
this fixed point in fact exists for the conformal window  
$\frakk{3}{2}N_c < N_f < 3N_c$.  
Our expansion clearly demands $N_f >> N_c$ and so  places the theory
firmly in the weakly coupled infra-red free regime;  nevertheless it is
tempting to compare our result Eq.~\sqcda\ with  perturbation theory for
a value of $N_f$ in the conformal  window.

\bigskip
\epsfysize= 3in
\centerline{\epsfbox{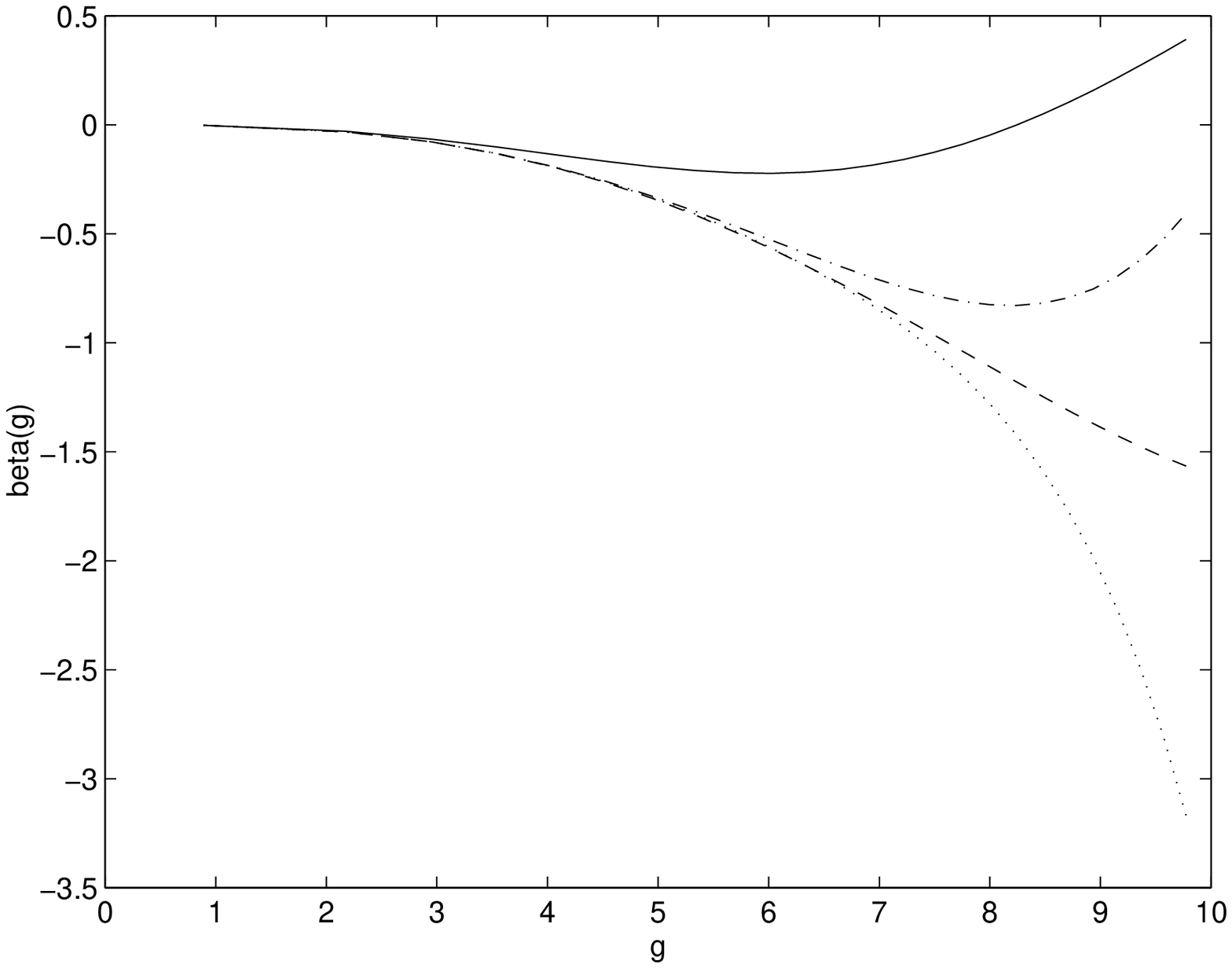}}
\in
{\it \noindent Fig.~7: Comparison between Eq.~\sqcda\ (solid line)
and the two, three and four loop approximations for $\beta_g$  
(dashed, dot-dashed and dotted
lines respectively), for $N_c = 3$ and  $N_f = 6$.}
\medskip
\out
In Fig.~7 we plot
$\beta_g$ against $g$ using $2\cdots 4$-loop  perturbation theory and
our result Eq.~\sqcda\ for $N_c = 3$ and  $N_f = 6$. In the case of
$\beta_g^{(4)}$ we have  set $\alpha = 0$; based on Ref.~\cjjc\ we would
anticipate that  $|\alpha|$ is $O(1)$, and the evolution is insensitive to the
value of  $\alpha$ in this region.  The IR fixed point occurs at
$g\approx 8$; of course this result is scheme dependent, but 
it is interesting that it is  substantially smaller  than that obtained in
either the two  or the three loop approximation; the four loop
approximation does not give  a fixed point at all (unless $\alpha$ is large and 
positive).  

\newsec{Discussion}

It is quite remarkable that the $O(1/N_f)$ corrections
to the SQCD $\beta_g$ depend only on simple integrals
involving $G(x)$. $G$ has a simple pole at $x=3/2$ and  
consequently $\beta_g$ has a logarithmic singularity at
$g^2 / (16\pi^2) = 3/2$ and a finite radius of convergence in $g$.
We have found that for values of $N_f$, $N_c$ corresponding to 
the conformal window $3N_c/2 < N_f <3N_c$, there indeed exists 
an infra-red fixed point in the gauge coupling evolution, though as we 
emphasised this regime is clearly outside the region of strict validity 
 of our approximation.

If we had some more terms in the $1/N_f$ expansion, we could try Pad\'e
or Borel-Pad\'e techniques and continue to small $N_f$. This might
permit us to extend the radius of convergence  mentioned above, and also
investigate more reliably the conformal window $3N_c/2 < N_f < 3N_c$.
Even the $O(1/N_f^2)$ contribution presents considerable technical
problems, however.  It is interesting to note that this calculation
would suffice to determine the unknown parameter $\alpha$ in Eq.~\loopb,
and consequently complete the derivation of $\beta_g^{(4)}$ carried out
in Ref.~\cjjc. Perhaps the critical methods of Ref~\honk\ could be
extended  to superfields and facilitate such calculations. It may also
be possible to determine $\alpha$\ref\JJS{I.~Jack, D.R.T.~Jones and
M.A.~Samuel, in  preparation} using Pad\'e
approximants\ref\EKS{J.~Ellis, M.~Karliner and M.A.~Samuel,
hep-ph/9612202}. It would be interesting to  see whether the finite
radius of convergence in $g$ which we noted above  persists at higher
orders in $1/N_f$. In fact, it is natural to speculate that the
$O(1/N_f^2)$ term, for instance, would depend on $G^2$, or some
convolution thereof.  

Since our calculations involve contributions from all orders in
perturbation theory, we should address the question of ambiguities in 
DRED\ref\amb{W.~Siegel, \plb94 (1980) 37\semi
L.V.~Avdeev, G.A.~Chochia and A.A.~Vladimirov, \plb105 (1981) 272.}
which can  potentially arise at higher orders. In our earlier paper on
the present topic\largen, we argued that the particular graphs we
consider,  consisting of bubble chains inserted onto simple lower order
graphs, are unambiguous; we also speculated that any DRED
ambiguities present in the theory  as a whole should be equivalent to
renormalisation scheme ambiguities and as  such can be subsumed into
coupling constant redefinitions.

\appendix{A}{Bubble sums}

In this appendix we give details of our bubble-sum calculations for the 
diagrams of Fig.~3. 
We do all calculations with zero external momentum, using 
\sic\ dimensional regularisation (with $d = 4-2\epsilon$)  and minimal
subtraction (DRED)\dred. By performing subtractions at the  level of the Feynman
integrals we completely separate  the calculation of the (subtracted)
Feynman integrals from  the details of the theory under consideration. 
It is convenient to redefine the  $d$-dimensional integration
measure so that 

\eqn\bubba{
\int \frakk{d^d k}{k^2 (k-p)^2} = \pi^2\frakk{1}{\epsilon}(p^2)^{-\epsilon}.
}
The diagrams we will require are shown in Figure~3.
Let us consider Fig.~3(B) first of all, as this is the simplest integral to
evaluate.  As explained in the main text, this diagram represents a sum 
over chains of bubbles of arbitrary length. 
The $n$-bubble diagram, before subtracting subdivergences, produces
a contribution
\eqn\gfunca{
\frakk{\kappa^n}{\epsilon^{n+1}} G(\epsilon)
{1-(n+1)\epsilon\over{n+1}}
\Gamma [1 + (n+1)\epsilon]\Gamma[1-(n+1)\epsilon]x^{(n+1)\epsilon}}
where
\eqn\gfunc{
G(\epsilon) = \frakk{\Gamma (2 -2\epsilon)}{\Gamma(2-\epsilon)
\Gamma(1-\epsilon)^2\Gamma(1+\epsilon)}} 
and $x = \mu^{-2}$, $\mu$ being the regulator mass. 
The parameter $\kappa$ subsumes any constant factors which will recur on
a bubble-by-bubble basis, including a factor of $(\lf)^{-1}$ for each 
bubble. Since this is an $(n+1)$-loop diagram, there remains an additional 
factor of $(\lf)^{-1}$ which we have included in the 
third columns of the Tables in the main text. Similar considerations will 
apply to Figs.~(A),(C), and (D).
We must now subtract diagrams with counterterm
insertions corresponding to
each divergent subdiagram. For the $n$-bubble contribution,
the divergent subdiagrams consist of subsets of $n-r+1$ bubbles, where 
$1\le r\le n$ (not necessarily forming a continuous chain). Such a subdiagram 
yields a counterterm of ${(-1)^{n-r}\kappa^{n-r+1}\over{\epsilon^{n-r+1}}}$, 
after subtracting all its own subdivergences. The remaining $(r-1)$-bubble 
diagram
gives a contribution as in Eq.~\gfunca, but with $n$ replaced by $r-1$. 
Taking into account a combinatorial factor of $\pmatrix{n\cr r-1\cr}$ for the
number of $(n-r+1)$-bubble subdiagrams, we find  
after subtracting all counterterm insertions that 
the $n$-bubble (i.e. $(n+1)$-loop) 
contribution to this diagram is given by the expression: 
\eqn\bubbb{
B_n = \frakk{\kappa^n}{\epsilon^{n+1}} G(\epsilon)
\sum_{r=1}^{n+1} r^{-1}(1-r\epsilon)
\Gamma (1 + r\epsilon)\Gamma(1-r\epsilon)\pmatrix{n\cr r-1\cr}(-1)^{r+1} 
x^{r\epsilon}.}
(We have also incorporated into this expression a factor of $(-1)^{L+1}$ for an
$L$-loop diagram, which derives from the $D$-algebra.) 

We now write\pmpp
\eqn\bubbc{
(1-r\epsilon)
\Gamma (1 + r\epsilon)\Gamma(1-r\epsilon)
x^{r\epsilon} = \sum_{j=0}^{\infty}L_j (r\epsilon)^j.}
Substituting in Eq.~\bubbb, and using the identity
\eqn\bubbd{\eqalign{\Delta_j = 
\sum_{r=1}^{n+1}r^{j-1}\pmatrix{n\cr r-1\cr}(-1)^r
&= 0 \quad\hbox{when}\quad j = 1,2,\cdots n\cr
&= -(n+1)^{-1} \quad\hbox{when}\quad j = 0\cr}} 
(which is proved in Appendix B) we find that the pole terms in $B_n$ are given 
by the expression
\eqn\bubbe{
B_n^{\rm{pole}} = \frakk{\kappa^n}{(n+1)\epsilon^{n+1}}
\sum_{i=0}^n G_i \epsilon^i
}
where we have written $G (\epsilon) = \sum G_n \epsilon^n$.    
The identity Eq.~\bubbd\ removes all the non-local (i.e. $\ln x$-dependent) 
counter-terms.  Now we want to sum over $n$. 
In a $\beta$-function or anomalous dimension calculation, 
the result will be given by the coefficient of the simple pole in $\epsilon$ 
in the quantity $\sum (n+1)B_n^{\rm{pole}}$, which is easily seen to give
\eqn\bubbf{
B(\kappa) =\sum_{n=0}^{\infty}G_n\kappa^n = G(\kappa).}

Next we consider Fig.~3(A). A complication as 
compared with the calculation for Fig.~3(B) is that there are now two types of 
divergent subdiagram; in addition to those consisting of subsets of bubbles,
there is also a divergent $(n+1)$-loop subdiagram formed by erasing the upper 
propagator in
Fig.~3(A). In fact this subdiagram is precisely the diagram Fig.~3(B), and 
therefore produces a counterterm $B_n^{\rm pole}$ as in Eq.~\bubbe. 
We therefore find that the $n$-bubble (or $(n+2)$-loop) contribution to the 
diagram is given by
\eqn\Aa{\eqalign{
A_n=&{\kappa^n\over{\epsilon^{n+2}}}\sum_{r=2}^{n+2}(-1)^{r-1}\left(\matrix{n\cr
r-2\cr}\right){H(\epsilon)^2\over{1-\epsilon}}{1\over{r(r-1)}}
{\Gamma(1-\epsilon+r\epsilon)\Gamma(1+\epsilon-r\epsilon)\Gamma(1+r\epsilon)
\over{\Gamma(1-2\epsilon+r\epsilon)}}x^{r\epsilon}
\cr
&+{\kappa^n\over{\epsilon^{n+2}}}{1\over{n+1}}\sum_{i=0}^n(G_i\epsilon^i)
{\Gamma(2-2\epsilon)\over{\Gamma(1-\epsilon)^2}}x^{\epsilon} ,\cr}}
where 
\eqn\Aaa{H(\epsilon)=\Gamma(2-\epsilon)G(\epsilon),}
 with $G(\epsilon)$ as in
Eq.~\gfunc. As for Fig.~3(B), we have also incorporated a factor $(-1)^{L+1}
=(-1)^{n+3}$ generated by the $D$-algebra.
We may now extend the summation in the second term to infinity, at the expense 
of introducing one extra simple pole term together with finite terms. We obtain,
up to finite terms, 
\eqn\Ab{\eqalign{
A_n=&{\kappa^n\over{\epsilon^{n+2}}}
\sum_{r=2}^{n+2}(-1)^{r-1}\left(\matrix{n\cr
r-2\cr}\right){H(\epsilon)^2\over{1-\epsilon}}{1\over{r(r-1)}}
{\Gamma(1-\epsilon+r\epsilon)\Gamma(1+\epsilon-r\epsilon)\Gamma(1+r\epsilon)
\over{\Gamma(1-2\epsilon+r\epsilon)}}x^{r\epsilon}
\cr
&+{\kappa^n\over{\epsilon^{n+2}}}{1\over{n+1}}
{H(\epsilon)^2\over{1-\epsilon}}
{\Gamma(1+\epsilon)\over{\Gamma(1-\epsilon)}}x^{\epsilon}
-{\kappa^n\over{\epsilon}}{1\over{n+1}}G_{n+1}\cr
=&{1\over{n+1}}{\kappa^n\over{\epsilon^{n+2}}}
\sum_{r=2}^{n+2}(-1)^{r-1}r^{-1}\left(\matrix{n+1\cr
r-1\cr}\right){H(\epsilon)^2\over{1-\epsilon}}
{\Gamma(1-\epsilon+r\epsilon)\Gamma(1+\epsilon-r\epsilon)\Gamma(1+r\epsilon)
\over{\Gamma(1-2\epsilon+r\epsilon)}}x^{r\epsilon}\cr
&+{\kappa^n\over{\epsilon^{n+2}}}{1\over{n+1}}
{H(\epsilon)^2\over{1-\epsilon}}
{\Gamma(1+\epsilon)\over{\Gamma(1-\epsilon)}}x^{\epsilon}
-{\kappa^n\over{\epsilon}}{1\over{n+1}}G_{n+1}.\cr}}
Rather remarkably, we now notice that the second term simply supplies the $r=1$
term in the summation of the first term. We therefore obtain
\eqn\Ac{\eqalign{
A_n=&{1\over{n+1}}{\kappa^n\over{\epsilon^{n+2}}}
\sum_{r=1}^{n+2}(-1)^{r-1}r^{-1}\left(\matrix{n+1\cr
r-1\cr}\right){H(\epsilon)^2\over{1-\epsilon}}
{\Gamma(1-\epsilon+r\epsilon)\Gamma(1+\epsilon-r\epsilon)\Gamma(1+r\epsilon)
\over{\Gamma(1-2\epsilon+r\epsilon)}}x^{r\epsilon}\cr
&-{\kappa^n\over{\epsilon}}{1\over{n+1}}G_{n+1}+\hbox{finite terms}.
\cr}}
In the same spirit as for Fig.~3(B), we write
\eqn\Ad{
{\Gamma(1-\epsilon+r\epsilon)\Gamma(1+\epsilon-r\epsilon)\Gamma(1+r\epsilon)
\over{\Gamma(1-2\epsilon+r\epsilon)}}x^{r\epsilon}=\sum_{j=0}^{\infty}
M_j(r\epsilon)^j.}
Substituting in Eq.~\Ac, and using Eq.~\bubbd, we find
\eqn\Ae{
A_n
={\kappa^n\over{\epsilon^{n+2}}}{H(\epsilon)^2\over{1-\epsilon}}{1\over
{(n+1)(n+2)}}M_0-{\kappa^n\over{\epsilon}}{1\over{n+1}}G_{n+1}
+\hbox{finite terms}.}
Clearly from Eq.~\Ad\ we have $M_0={\Gamma(1-\epsilon)\Gamma(1+\epsilon)
\over{\Gamma(1-2\epsilon)}}$, and so, from Eqs.~\gfunc, \Aaa, we find
\eqn\Af{\eqalign{
A_n
=&{\kappa^n\over{\epsilon^{n+2}}}(1-2\epsilon){1\over{(n+1)(n+2)}}
G(\epsilon)\cr
&-{\kappa^n\over{\epsilon}}{1\over{n+1}}G_{n+1}+\hbox{finite terms}.
\cr}}
As in the case of Fig.~3(B), the contribution to a $\beta$-function or 
anomalous dimension will be given by the coefficient of the simple pole in
$\epsilon$ in the quantity $\sum(n+2)A_n$, giving
\eqn\Ag{\eqalign{
A(\kappa)=&\sum_{n=0}^{\infty}{1\over{n+1}}(G_{n+1}-2G_n)\kappa^n
-\sum_{n=0}^{\infty}{n+2\over{n+1}}G_{n+1}\kappa^n\cr
=&-\sum_{n=0}^{\infty}\left[G_{n+1}+{2\over{n+1}}G_n\right]\kappa^n ,\cr}}
which may be rewritten in the compact form 
\eqn\Aga{
A(\kappa)
=-{1\over{\kappa}}\left[G(\kappa)-1+2\int_0^{\kappa}G(x)\,dx\right].}
The calculation of $C(\kappa)$ parallels very closely that of $A(\kappa)$, and 
yields 
\eqn\Ah{
C(\kappa) = - 2\kappa^{-1}\left[G(\kappa) -1 + \int_0^\kappa (1+2x)G(x)\,dx
\right],}
The computation of Fig.~3(D) is somewhat
different, however, and we shall explain it in detail. The $n$-bubble diagram
together with its counterterms (in this case simply bubble-chains, as for 
Fig.~3(B)) gives
\eqn\Ai{
D_n=-(n+1){\kappa^n\over{\epsilon^{n+1}}}\sum_{r=3}^{n+3}{(-1)^{r-1}\over r}
\left(\matrix{n\cr r-3\cr}\right)G(\epsilon)
\Gamma(1-r\epsilon)\Gamma(1+r\epsilon)x^{r\epsilon}.} 
The factor of $(n+1)$ represents the
$(n+1)$ ways of distributing $n$ bubbles in Fig.~3(D).
Now we write
\eqn\Aj{\Gamma(1-r\epsilon)\Gamma(1+r\epsilon)x^{r\epsilon}
=\sum_{j=0}^{\infty}P_j
(r\epsilon)^j.}
It is easy to see from Eq.~\bubbd\ that 
\eqn\Ak{
\sum_{r=3}^{n+3}(-1)^rr^{j-1}\left(\matrix{n\cr r-3\cr}\right)=0}
for $j\ge 1$. This ensures the cancellation of non-local terms and leaves
only the $j=0$ term. Using the result 
\eqn\Al{
\sum_{r=3}^{n+3}{(-1)^r\over r}\left(\matrix{n\cr r-3\cr}\right)=-
{2\over{(n+1)(n+2)(n+3)}},}
which is proved in Appendix B, together with $P_0=1$, we have
\eqn\Am{
D_n=-{\kappa^n\over{\epsilon^{n+1}}}G(\epsilon){2\over{(n+2)(n+3)}}.}
The contribution to a $\beta$-function or anomalous dimension is then given by
the simple pole in $\sum(n+3)D_n$, yielding
\eqn\An{
D(\kappa)=-2\sum_{n=0}^{\infty}{G_n\over{n+2}}\kappa^n,}
which may be rewritten
\eqn\Ana{
D(\kappa)=-{2\over{\kappa^2}}\int_0^{\kappa}xG(x)\, dx.}

\appendix{B} {Summation identities}
In this appendix we prove the identities Eqs.~\bubbd\ and \Al, 
which play a crucial r\^ole in disposing of non-local contributions and deriving
the final bubble-sum results. Eq.~\bubbd\ may be proved starting from 
\eqn\Ba{
(1-x)^n=\sum_{r=0}^n(-1)^rx^r\left(\matrix{n\cr r\cr}\right).}
Differentiating Eq.~\Ba\ $l$ times (where $l\le n-1)$, and setting $x=1$, 
we obtain
\eqn\Bc{
0=\sum_{r=0}^n(-1)^rr(r-1)\ldots(r-l+1)\left(\matrix{n\cr r\cr}\right).}
Clearly, by taking linear combinations of Eq.~\Bc\ with different values of $l$,
we may obtain
\eqn\Bf{
0=\sum_{r=0}^n(-1)^rr^{j-1}\left(\matrix{n\cr r\cr}\right)}
for $1\le j\le n$. On the other hand, by integrating Eq.~\Ba\ we obtain
\eqn\Bg{
{1\over {n+1}}=\sum_{r=0}^n(-1)^r(r+1)^{-1}\left(\matrix{n\cr r\cr}\right).}
Upon changing the summation range, Eqs.~\Bf\ and \Bg\ are clearly equivalent
to Eq.~\bubbd, namely
\eqn\Bh{\eqalign{\Delta_j =
\sum_{r=1}^{n+1}r^{j-1}\pmatrix{n\cr r-1\cr}(-1)^r
&= 0 \quad\hbox{when}\quad j = 1,2,\cdots n\cr
&= -(n+1)^{-1} \quad\hbox{when}\quad j = 0.\cr}}
Upon integrating the identity
\eqn\Bi{
x^2(1-x)^n=\sum_{r=0}^n(-1)^r\pmatrix{n\cr r\cr}x^{r+2},}
and then setting $x=1$, we obtain 
\eqn\Bi{ 
\sum_{r=0}^{n}{(-1)^r\over r+3}\left(\matrix{n\cr r\cr}\right)=
{2\over{(n+1)(n+2)(n+3)}},}
which is equivalent to Eq.~\Al.
\bigskip\centerline{{\bf Acknowledgements}}\nobreak

PF was supported by a  scholarship from JNICT, and CGN by a PPARC
studentship. We thank John Gracey for
several useful  discussions. 

\listrefs 

\bye